\documentclass[twocolumn]{aastex631}

\usepackage{amssymb}
\usepackage{amsmath}
\usepackage{amssymb}
\usepackage{empheq}
\usepackage{mathrsfs}
\usepackage{wasysym} 
\usepackage{fourier-orns}
\usepackage{adforn}

\usepackage{etoolbox}
\usepackage{lmodern}

\newcommand{\be}{\begin{align}}
\newcommand{\ee}{\end{align}}

\newcommand{\appropto}{\mathrel{\vcenter{
  \offinterlineskip\halign{\hfil$##$\cr
    \propto\cr\noalign{\kern2pt}\sim\cr\noalign{\kern-2pt}}}}}

\begin{document} 

\title{\bf Suppression of Resonant Overstability at Sharp Migration Gradients}

\author[0000-0002-7094-7908]{Konstantin Batygin}
\affiliation{Division of Geological and Planetary Sciences, California Institute of Technology, Pasadena, CA 91125}

\author[0000-0002-2478-3084]{Ian R. Brunton}
\affiliation{Division of Geological and Planetary Sciences, California Institute of Technology, Pasadena, CA 91125}

\author[0000-0001-8476-7687]{Alessandro Morbidelli}
\affiliation{Laboratoire Lagrange, Universit\'e C\^ote d'Azur, Observatoire de la C\^ote d'Azur, CNRS, CS 34229, F-06304 Nice, France}

\begin{abstract}
Mean-motion resonances are expected to frequently arise at the inner edges of protoplanetary disks, where planet-disk interactions facilitate large-scale orbital convergence. Under certain conditions, however, the same dissipative forces that promote resonant capture can drive resonant librations overstable, ultimately breaking commensurabilities. Here we examine the onset of overstability near disk torque reversals and show that it can be subdued when the transition is sufficiently sharp. Adopting the dissipative circular restricted three-body problem as a paradigm, we present a WKB-style analysis that reduces the resonant dynamics to a damped, driven harmonic oscillator. Within this framework, we obtain an effective frictional term that is proportional to the local migration-rate gradient, parameterized by a dimensionless coefficient $\beta$ that encodes the steepness of the local torque reversal. Our analytical theory predicts that overstability is quenched once $\beta \gtrsim \tau_a/\tau_e$, where $\tau_a$ and $\tau_e$ denote the characteristic disk-driven evolution timescales of semi-major axis and eccentricity. We verify and refine our analytic results with direct $N$-body integrations. Simple estimates based on conventional type-I scalings suggest that the competition between overstability and its mitigation at disk inner edges is a borderline outcome that is sensitive to the detailed structure of planet-disk interactions.
\end{abstract}

\keywords{Planetary dynamics}

\section{Introduction}

Geometrically thin, nearly-Keplerian disks rank among the most common structures within the known universe, occupying an enormous span of physical regimes. They range from tenuous circumplanetary rings to circumstellar (protoplanetary) nebulae, extending to the gaseous disks that encircle supermassive black holes in active galactic nuclei, and even to populations of stars within inner regions of galaxies themselves. A shared process that operates across this entire hierarchy of astrophysical environments is that massive bodies embedded within disks inevitably excite gravitational perturbations. The disks respond, generating a time irreversible exchange of energy and angular momentum that drives orbital migration \citep{1980ApJ...241..425G,1986ApJ...309..846L, 1997Icar..126..261W}. In systems where multiple objects experience migration simultaneously, convergent radial drift of neighboring orbits naturally leads to the emergence of mean-motion resonances \citep{1975JApMM..39..594N, 1976ARA&A..14..215P,1982CeMec..27....3H}.

Instances of observed resonant structures are numerous. The Jovian (Galilean) and Saturnian satellites provide classical solar system examples \citep{1954MNRAS.114..232R,1965MNRAS.130..159G,1968MNRAS.141..363D,1972MNRAS.160..169S,1973MNRAS.165..305G}, while the census of extrasolar planets hosts an ever-increasing count of resonant pairs and chains \citep{2002ApJ...567..596L,2016Natur.533..509M,2023Natur.623..932L,2016MNRAS.455L.104G,2023AJ....165...33D}. Their prevalence is generally attributed to the existence of convergence zones or ``migration traps’’ within disks, where orbital commensurabilities are established efficiently. A canonical example in circumplanetary and protoplanetary nebulae is the sharp surface-density drop at the magnetospheric cavity: while global disk torques typically drive inward orbital decay \citep{2002ApJ...565.1257T}, the steep density transition at the inner edge greatly enhances outward-directed corotation torques, leading to a reversal of migration direction \citep{2006ApJ...642..478M}. As a result, radially drifting planets or satellites naturally accumulate at disk inner edges and become locked into resonant configurations. Similar behavior emerges in active galactic nuclei, where thermal torques produce analogous convergence zones \citep{2024MNRAS.530.2114G} and shepherd stars or compact objects into resonant configurations that may precede merger events \citep{2025arXiv251012895E}.

Curiously, the same dissipative processes that enable resonance capture can, under certain circumstances, destabilize it. In their seminal analysis, \citet{GS14} re-examined the circular restricted three-body problem under dissipation, and showed that resonant librations can become overstable, ultimately breaking the commensurability. \citet{2015ApJ...810..119D} extended this result to the full (unrestricted) three body problem and demonstrated that overstability occurs only when the inner body is significantly less massive than the outer. \citet{2015A&A...579A.128D} presented a complementary treatment that quantified the onset of overstability within a general Hamiltonian framework, while \citet{2017MNRAS.468.3223X} confirmed that the effect persists for resonances of order exceeding unity. Thus, beyond being a technical peculiarity of dissipative resonant dynamics, overstability of resonant librations constitutes a general process that almost certainly plays a meaningful role in sculpting the observed census of resonant planetary systems. As one concrete example set within our own Solar System, the onset of overstability between Europa and Ganymede dictates the inside-out order of assembly of the Laplace resonance \citep{2020ApJ...894..143B,2026arXiv260100786Y}, while the overstability of Io’s interior resonances during the inward transport of Jovian inner moons (e.g., Amalthea and Thebe) places tight constraints on the thermodynamic state of the primordial circum-Jovian nebula \citep{2025ApJ...991...15B,2025ApJ...990L..11B}.

At its core, the fundamental mechanism that underpins overstability lies in the coupling between eccentricity evolution and migration. That is, the central refinement over earlier treatments identified by \citet{GS14} was to include the eccentricity dependence of the migration rate within the perturbation equations. In this work, we examine the contrasting consequences of allowing the migration rate to vary not only with eccentricity but also with semi-major axis itself, as expected near disk transitions. Through analytic and numerical calculations, we show that sufficiently sharp torque reversals can suppress the onset of overstability entirely.

The remainder of this paper is structured as follows. In Section~2, we lay out the analytic formulation of the problem and show, using a WKB-type treatment, that the stability of the dissipative circular restricted three-body problem reduces to that of a damped, driven harmonic oscillator. Section~3 presents numerical simulations that verify and sharpen the analytic expectations. In Section~4, we discuss the implications of overstability suppression at disk transitions.

\section{Analytic Theory}

A central objective of this calculation is to understand, from analytic grounds, the conditions under which dissipative librations in a first-order mean-motion resonance transition from being overstable to having their amplitude be suppressed. As already alluded to above, prior work has established two key requirements for overstability to arise in the first place. First, the mass ratio between the inner and outer bodies must be sufficiently small. A second, somewhat more subtle constraint is that the resonant equilibrium must lie sufficiently close to nominal commensurability. This prerequisite places a condition on the relationship between the mass of the perturber and ratio of the disk-driven evolution rates of eccentricity and semi-major-axis. With these ingredients in place, our strategy is to construct the problem so that overstable librations serve as the baseline state. We therefore work within the paradigm of the circular restricted three-body problem\footnote{The circular restricted approximation is less limiting than it may appear. To leading order in eccentricity, the unrestricted problem admits an equivalent integrable resonant Hamiltonian (see, e.g., \citealt{2013A&A...556A..28B}). A complementary discussion of the connection between the restricted and full formulations is given in the recent work of \citet{2025ApJ...986...11T}.} with an exterior perturber, and make a series of simplifying approximations so that the derivation remains as transparent as possible.

\begin{figure}
\includegraphics[width=\columnwidth]{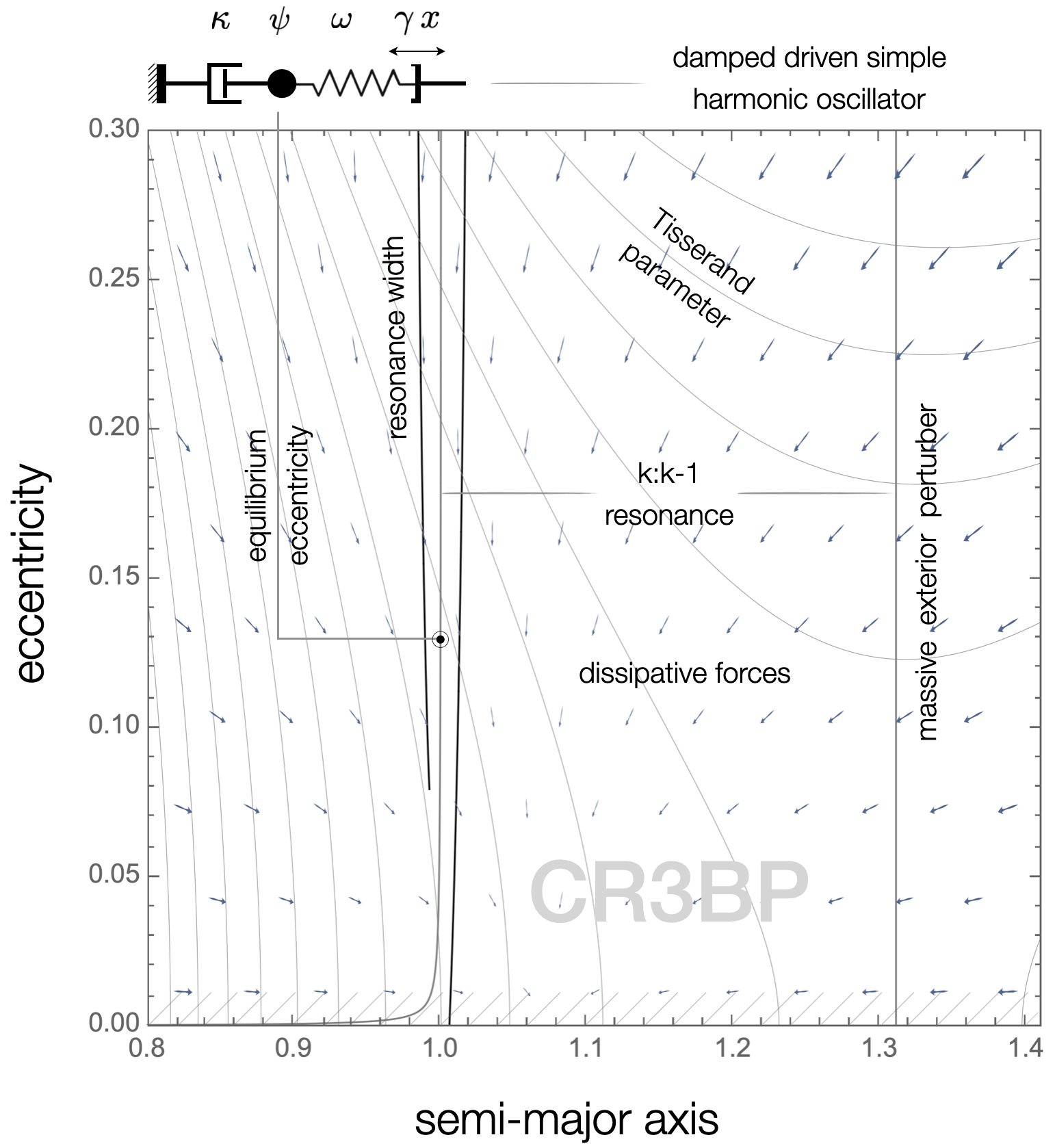}
\centering
\caption{Schematic of the dissipative circular restricted three-body problem. A massless test particle resides in an interior $k:k-1$ resonance (here, taken to be $k=3$) with a circular exterior perturber, and evolves subject to disk-driven evolution of $a$ and $e$. The resonance width (corresponding to a perturber of mass $m' =10^{-4}\,M$) is illustrated with bold black curves while the resonant equilibrium is shown with a gray line at a semi-major axis of approximately unity. Background gray contours denote the Tisserand parameter. Disk-driven interactions facilitate eccentricity damping and semi-major axis evolution that depends on both $e$ and $a$. This effect is depicted as a vector-field on the figure (with the torque reversal steepness parameter set to $\beta = 8$. Higher $e$ introduces an $e^{2}$ decay component of $a$. Simultaneously, the migration rate also varies with $a$ itself, across a threshold of orbital convergence (a ``trap"), such that outward migration is faster at smaller $a$. Linearized motion about the equilibrium eccentricity reduces to a damped, driven harmonic oscillator for the resonant phase $\psi$, with the effective damping set by the local migration-rate gradient. The hatched low-eccentricity region denotes a regime where the slow-precession assumption of our analytic model breaks down.}
\label{fig:CRTBP}
\end{figure}

A schematic of the physical setup is shown in Figure \ref{fig:CRTBP}. In the conservative planar circular restricted three-body problem, motion is constrained by the Jacobi integral, whose approximate osculating-element form may be written as the Tisserand parameter, $\mathcal{T}=a'/a+2\sqrt{(a/a')\left(1-e^2\right)}$, where the primed quantities refer to the exterior perturber \citep{1999ssd..book.....M}. The thin gray background curves in Figure \ref{fig:CRTBP} denote contours of constant \(\mathcal{T}\) and are included as a guide to the local conservative phase-space geometry of the resonance. Dissipative effects, represented schematically by the vector field, drive a slow secular drift across these contours until the resonant fixed point is reached. Small-amplitude librations about this equilibrium are then approximately tangent to constant-\(\mathcal{T}\) curves to leading order. Accordingly, in this section we first establish the resonant equilibrium and then examine the local behavior of librations about it. The goal is not to re-derive every aspect of the theory outlined by \citet{GS14, 2015ApJ...810..119D,2015A&A...579A.128D} and others, but rather to extend their framework to include semi-major-axis dependence of migration near torque reversals, and to determine analytically when this effect quenches overstability.

\subsection{Resonant Equilibrium} \label{RE}

To first order in eccentricity and inclination, the disturbing function associated with an interior first-order resonance in the circular restricted problem has the form  
\begin{align}
\mathcal{R} \;=\; \frac{G \,m'}{a'}\, f\, e\,\cos\phi,
\end{align}
where the resonant argument is
\begin{align}
\phi = k\,\lambda' - (k-1)\,\lambda - \varpi.
\label{eqn:phi}
\end{align}
Here, Keplerian orbital elements take on their customary meanings, the integer $k>1$ denotes the resonance index, and \(f\) is the usual (order unity and negative) combination of Laplace-coefficients, dependent only on the semi-major-axis ratio \(a/a'\).

The proximity to exact resonance is quantified by the detuning parameter  
\begin{align}
\Delta = \frac{k\, n' - (k-1)\,n}{n}.
\end{align}
In what follows, we seek the resonant fixed point by solving sequentially for the three unknowns: the equilibrium eccentricity $e_{\rm eq}$, the equilibrium resonant phase $\phi_{\rm eq}$, and finally the equilibrium detuning, $\Delta_{\rm eq}$ (equivalently the equilibrium semi-major axis). Concretely, we first impose $\dot{\phi}=0$ to relate the detuning to the apsidal precession rate; this immediately yields $e_{\rm eq}$ once the stable branch of $\phi$ is identified. We then enforce $\dot{e}=0$ to determine $\phi_{\rm eq}$, and close the solution by requiring $\dot{a}=0$ to obtain $\Delta_{\rm eq}$.

\medskip

\paragraph{Equilibrium eccentricity $e_{\rm eq}$\, } \ \  Stationarity of the resonant argument demands  
\begin{align}
\dot{\phi} = 0 \qquad\Rightarrow\qquad 
\dot{\varpi} = k n' - (k-1)n = n\,\Delta.
\label{eqn4}
\end{align}
The rate of apsidal precession follows from Lagrange’s equation,
\begin{align}
\left( \frac{d\varpi}{dt} \right)_{\mathcal{R}}
= \frac{1}{n a^2 e}\frac{\partial \mathcal{R}}{\partial e}
= \frac{G m'}{a'}\,\frac{f\cos\phi}{n a^2 e}.
\label{eqn5}
\end{align}

From a simple geometric argument (see Appendix), it is straightforward to show that for the interior resonant periodic orbit, apsidal precession is retrograde, and $\Delta < 0$. Since \(f\) is also negative, the above equation dictates that the resonant argument must approximately be $\phi\sim0$, such that $\cos\,\phi\sim1$. Approximating the semi-major axis ratio as $a/a' \simeq ((k-1)/k)^{2/3}$ and solving for \(e\) yields the resonant equilibrium eccentricity
\begin{align}
e_{\rm eq}
= \frac{f}{\Delta}\,\frac{m'}{M}\left(\frac{k-1}{k}\right)^{2/3}.
\label{eeq}
\end{align}

\medskip

\paragraph{Equilibrium argument $\phi_{\rm eq}$\, } \ \ Adopting linear damping of the form  
\begin{align}
\left(\frac{de}{dt}\right)_{\rm diss} = -\frac{e}{\tau_e},
\label{eqn:dedt}
\end{align}
the Lagrange equation for eccentricity gives the following stationary condition:
\begin{align}
-\frac{e}{\tau_e}
= -\left(\frac{de}{dt}\right)_{\mathcal{R}}
= \frac{1}{n a^2 e}\frac{\partial\mathcal{R}}{\partial\varpi}
= \frac{G m'}{a'}\,\frac{f \sin\phi}{n a^2}.
\label{eqn:ecc}
\end{align}
Substituting the solution (\ref{eeq}) for the eccentricity yields the equilibrium resonant angle
\begin{align}
\sin\phi_{\rm eq}
= -\frac{1}{n\,\Delta\,\tau_e}.
\end{align}
In the practically relevant weak-friction limit, $|\Delta\tau_e| \gg 1$ and $\Delta<0,$ so \(\phi_{\rm eq}\) is small and positive.

\medskip

\paragraph{Equilibrium detuning $\Delta_{\rm eq}$\, }\ \ To close the equilibrium solution, we impose stationarity of the semi-major axis. Because we are interested in dynamics near a migration trap, we allow the disk-driven migration rate to depend both on eccentricity and on orbital frequency. Over the narrow resonance width, we may linearize the frequency dependence:
\begin{align}
\left(\frac{1}{a}\frac{da}{dt}\right)_{\rm diss}
= \frac{1}{\tau_a}\left[1 + \beta\,\frac{\delta n}{[n]}\right]
- \frac{2 e^2}{\tau_e},
\label{dadtdiss}
\end{align}
where \(\beta>0\) parameterizes the steepness of the torque gradient near the trap, and \(\delta n = n - [n]\), with \([n]\) corresponding to the \(\dot{\phi}=0\) state. The corresponding disk-driven evolution in the ($a,e$) plane, combining eccentricity damping with migration whose sign and magnitude vary across the torque reversal (evaluated from equations \ref{dadtdiss} and \ref{eqn:dedt}), is illustrated in Figure~\ref{fig:CRTBP} as a vector field.


Invoking Lagrange's equations one last time, we compute the resonant contribution to the semi-major axis evolution:
\begin{align}
&\left(\frac{1}{a}\frac{da}{dt}\right)_{\mathcal{R}} = \frac{2}{n\,a^2} \frac{\partial\mathcal{R}}{\partial\lambda} = \frac{2\,(k-1)}{n\, a^2}\frac{G m'}{a'} f\,e\sin\phi.
\label{dadtR}
\end{align}
Substituting expression obtained above for \(e_{\rm eq}\) and \(\sin\phi_{\rm eq}\) into the stationary condition for the semi-major axis:
\begin{align}
&\left(\frac{1}{a}\frac{da}{dt}\right)_{\rm diss} + \left(\frac{1}{a}\frac{da}{dt}\right)_{\mathcal{R}} = 0,
\end{align}
we solve for $\Delta$, to obtain the equilibrium detuning:
\begin{align}
\Delta_{\rm eq}
= f\,\frac{m'}{M}\left(\frac{k-1}{k}\right)^{2/3}
\sqrt{\frac{2\,k\,\tau_a}{\tau_e}}.
\end{align}

\subsection{Linearized Oscillations}

We now consider small-amplitude librations about the equilibrium. To this end, we define small quantities
\begin{align}
x \equiv e - e_{\rm eq}, \qquad \psi \equiv \phi - \phi_{\rm eq}.
\label{xpsi}
\end{align}
Furthermore, we adopt the ansatz:
\begin{align}
\psi(t)=\mathcal{A}(t)\cos(\nu t),
\end{align}
with the “slow-amplitude” condition \(|\dot{\mathcal{A}}|\ll\nu\mathcal{A}\), and write the variational equations for the eccentricity and the resonant argument.

\medskip

\paragraph{Evolution of $x$\, } \ \ Linearizing the full eccentricity equation about the fixed point gives:
\begin{align}
\frac{dx}{dt} = - \frac{x}{\tau_e}  -f\,n\,\frac{m'}{M}\left(\frac{k-1}{k}\right)^{2/3}\,\psi,
\label{eqn:dxdt}
\end{align}
Strictly speaking, the linearization also generates an additional term -- not present in the above equation -- arising from the variation of the factor $f/(n\,a^2)$ in equation (\ref{eqn:ecc}). This contribution, however, is  proportional to $\sin\phi_{\rm eq}\,(\delta a/a)$, and in the weak-friction regime, where $\phi_{\rm eq}\sim0$, this term is sub-leading. For the same reason, we set $\cos\phi_{\rm eq}\simeq 1$. Note also that, in writing the coefficient of $\psi$, we have expressed the resonant semi-major-axis ratio as $a/a'\simeq ((k-1)/k)^{2/3}$.

Substituting the ansatz for \(\psi\) gives the solution
\begin{align}
x(t) = \eta\,\mathcal{A}(t)\left[ \cos(\nu t) + \nu\tau_e\sin(\nu t) \right] + \sigma \,e^{-t/\tau_e}.
\label{eqn:xt}
\end{align}
The pre-factor in the above expression has the form:
\begin{align}
\eta\simeq -f\,n\,\tau_e\,\frac{m'}{M} \left(\frac{1}{\tau_e\,\omega}\right)^2 \left(\frac{k-1}{k}\right)^{2/3},
\end{align}
where we have approximated $(1+\nu^2\,\tau_e^2)\simeq(\omega\,\tau_e)^2$, with $\omega$ being the nominal libration frequency (note also that in the regime of interest, \(\tau_e\,\omega\gg1\)).

The integration constant \(\sigma\) (which encapsulates all dependence on initial conditions), decays away on the eccentricity-damping timescale and is henceforth neglected. Importantly, the presence of both sine and cosine terms in equation (\ref{eqn:xt}) indicates a slight phase offset between \(x\) and \(\psi\). This mismatch is a key ingredient that acts as the drive for overstability.

\medskip

\paragraph{Evolution of $\psi$\, } \ \ Near the resonant fixed point, the variables $x$ and $\psi$ exhibit coupled oscillations, somewhat analogously to a coordinate-momentum pair in the Hamiltonian description. Accordingly, one may describe their dynamics through a pair of first-order equations: equation~(\ref{eqn:dxdt}) governs the dissipative response of the eccentricity perturbation, while the phase evolution is tied to the resonant detuning. It is nevertheless convenient to describe the evolution in terms of a single second-order equation for $\psi$, because this makes the oscillator structure explicit, elucidating how eccentricity fluctuations drive the phase dynamics.

To obtain the phase equation, we differentiate the resonant angle twice with respect to time and express the result in terms of the departure from exact commensurability. Neglecting $\ddot{\varpi}$ in favor of $\dot{n}$ (the same approximation used earlier and appropriate in the sufficiently small-$m'/M$ regime where the baseline state is overstable; see Appendix B), we write:
\begin{align}
\ddot{\phi}\simeq -(k-1)\dot{n} = \frac{3\,(k-1)\,n}{2} \frac{\dot{a}}{a}.
\end{align}
At the order retained here, we treat $x$, $\psi$, and $\delta n$ as first-order quantities, while $e_{\rm eq}$, $\phi_{\rm eq}$, and $\Delta_{\rm eq}$ are held fixed at their equilibrium values. More specifically, in varying the semi-major axis evolution, we retain the $(\beta/\tau_a)(\delta n/[n])$ and $-(4\,e_{\rm eq}/\tau_e)\,x$ terms from the dissipative contribution\footnote{Because $[n]$ denotes the local $\dot{\phi}=0$ state, the first derivative of equation (\ref{eqn:phi}) gives $\dot{\psi}=\dot{\phi}\approx-(k-1)\,\delta\,n$. This is the origin of the term proportional to $\dot{\psi}$ in equation (\ref{SHO}).} (equation \ref{dadtdiss}) and terms proportional to $e_{\rm eq}\psi\cos\phi_{\rm eq}$ and $x\sin\phi_{\rm eq}$ from the resonant contribution, while neglecting the variation of the prefactor $f/(na^2)$, as above.

Then, noting that $\ddot{\psi}=\ddot{\phi}$ from equation (\ref{xpsi}), and writing the expression explicitly in terms of $x$ and $\psi$ we obtain:
\begin{align}
\ddot{\psi} = -\omega^{2}\psi - \kappa\,\dot{\psi} - \gamma\,x + \mathcal{O}(x^2,\,x\psi),
\label{SHO}
\end{align}
where the coefficients are:
\begin{align}
&\omega = n \left( \frac{3\,|f|\, (k-1)^{8/3}}{k^{7/6}}\,\frac{m'}{M}\sqrt{\frac{\tau_e}{2\,\tau_a}} \right)^{1/2}, \nonumber \\
&\kappa = \frac{3\beta}{2\tau_a}, \qquad \gamma = 3n\left(\frac{k^2-1}{k}\right)\sqrt{\frac{k}{2\tau_a\tau_e}}.
\end{align}

Clearly, to leading order in small quantities, the evolution of $\psi$ simplifies to the familiar paradigm of a damped, driven, harmonic oscillator (as it should). Moreover, the physical meaning of the coefficients is intelligible: while \(\omega\) represents the natural resonant libration frequency and \(\gamma\) parameterizes the degree of coupling between eccentricity and resonant angle, \(\kappa\) is the dissipative coefficient arising from the torque gradient.

Substituting our ansatz for $\psi$ and solution for $x$ into the expression for \(\ddot{\psi}\) produces terms proportional to \(\mathcal{A}\), \(\dot{\mathcal{A}}\), and \(\ddot{\mathcal{A}}\). As in the usual WKB approximation, we neglect the \(\ddot{\mathcal{A}}\) term because \(\mathcal{A}\) varies slowly compared to the oscillation timescale and \(\ddot{\mathcal{A}}\) is slower still. Finally, equating the terms proportional to \(\cos(\nu t)\) and \(\sin(\nu t)\) separately yields conditions for the frequency and the amplitude:
\begin{align}
\nu^{2} = \omega^{2} + \gamma\,\eta + \kappa\,\frac{\dot{\mathcal{A}}}{\mathcal{A}}, \nonumber \\
\dot{\mathcal{A}} = -\frac{1}{2}\bigl(\kappa - \gamma\eta\tau_e\bigr)\mathcal{A}.
\end{align}
In practice, we expect frequency corrections to be small, such that $\nu\sim\omega$ to leading order. The derivative of the amplitude, on the other hand, can be either positive or negative.

\paragraph{Suppression of Overstability \, } \ \  In the limit of free migration (\(\kappa\rightarrow 0\)), the amplitude grows exponentially on a timescale
\begin{align}
\tau_{\rm grow}
= \frac{2(k-1)^2}{k^2-1}\,\tau_e
\sim \tau_e,
\end{align}
reproducing the Goldreich--Schlichting result. In a migration trap, however, the sign of \(\dot{\mathcal{A}}\) reverses if $\kappa>\gamma\,\eta\,\tau_e$, which gives the criterion
\begin{align}
\beta>\frac{2}{3}\frac{k+1}{k-1}\frac{\tau_a}{\tau_e}
\sim \frac{\tau_a}{\tau_e}.
\label{betacrit}
\end{align}
In other words, the gradient of torque reversal must be sharper than the ratio of extrinsically-forced semi-major axis to eccentricity evolution timescales for overstability to be suppressed.


Although the foregoing criterion is transparent and captures the essential physics, it is useful to note that the approximation $|\ddot{\varpi}| \ll |\dot n|$ removes the explicit dependence on the perturber-to-central mass ratio, $m'/M$, that enters through the resonant contribution to apsidal precession. This simplification is convenient for obtaining an analytic suppression criterion, but it is not fundamental. In Appendix~\ref{app:B}, we show that once $\ddot{\varpi}$ is retained, the linear theory acquires a corresponding mass-dependent correction, thereby recovering the expected connection to the threshold identified by \citet{GS14}. At the same time, this refinement does not alter the basic structure of the result: differential migration still suppresses overstability once it is sufficiently strong compared with the delayed eccentricity forcing. We therefore proceed to numerical experiments, which confirm the analytic picture and sharpen the location of the stability boundary.

\section{Numerical Simulations}

\begin{figure}
\includegraphics[width=\columnwidth]{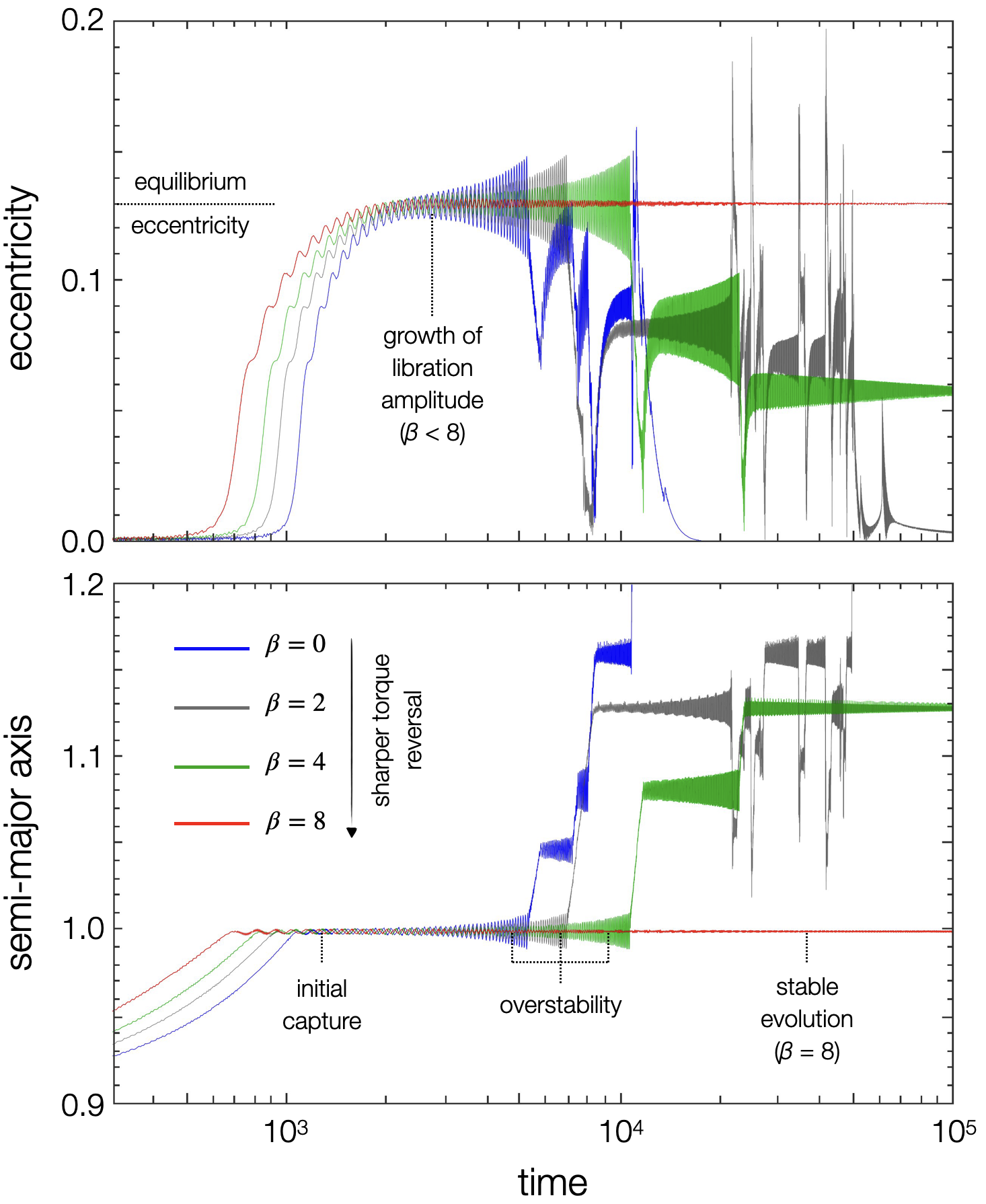}
\centering
\caption{Numerical verification of overstability quenching. Time-series of eccentricity (top) and semi-major axis (bottom) are shown for integrations of the dissipative circular restricted three-body problem with $m'=10^{-4}M_\star$ and $\tau_a/\tau_e=10$, for a range of torque-gradient parameters $\beta$. In all runs, outward migration first produces capture into the interior 3:2 resonance and growth of $e$ toward the equilibrium value. For $\beta=0$ the libration amplitude subsequently widens and the resonance is disrupted, while intermediate values ($\beta=2,4$) weaken but do not eliminate the overstable growth. For a sufficiently sharp torque reversal ($\beta=8$), libration amplitude decays and the evolution is stable. Notably, quenching of overstability occurs at a somewhat smaller torque gradient than predicted by the analytic estimate of $\beta\gtrsim13$ (see equation \ref{betacrit}).}
\label{fig:timeseries}
\end{figure}

To verify the analytic expectations developed above, we carried out a suite of numerical experiments using the \texttt{REBOUND} $N$-body integrator \citep{ReinLiu2012} augmented with the \texttt{modify\_orbits\_forces} prescription from \texttt{REBOUNDx} \citep{Tamayo2020}. The integrations were performed with the \texttt{mercurius} scheme. Taking advantage of the scale-free nature of Newtonian gravity, we worked in dimensionless units in which the semi-major axis ($\tilde{a}$) and orbital period ($\tilde{P}$) of the nominal (Keplerian) interior 3:2 commensurability are taken to be unity. In these units, the exterior perturber of mass\footnote{We note that a dimensionless mass ratio of order $10^{-4}$ is comparable to the total satellite-to-planet mass fraction of the Solar System’s giant planets, and lies within an order of magnitude of the typical cumulative planet-to-star mass ratios inferred for compact Kepler multi-planet systems, although the latter span a relatively broad range.} $m' = 10^{-4} M_\star$ resides at $\tilde{a}' = (3/2)^{2/3}$, while the massless test particle is initialized on a circular, coplanar orbit at $\tilde{a}=0.9$, interior to exact resonance. The timestep was set to $\Delta t = 10^{-2}\,\tilde{P}$, and the imposed migration caused the test particle to drift outward toward the 3:2 commensurability.

The eccentricity of the test particle was damped with a fixed timescale $\tau_e = 1000\,\tilde{P}$, while semi-major axis evolution was controlled through a migration timescale that depended on the instantaneous mean motion, as prescribed in the previous section (equation \ref{dadtdiss}). The nominal migration rate was set to $\tau_a = 10\,\tau_e$, while $\beta$ was retained as a free parameter, determining the steepness of the local torque gradient. We explored values $\beta = 0,\,1,\,2,\,4,\,8,$ and $16$, ranging from a spatially uniform migration rate to gradients comparable to those expected near a sharp disk inner edge. Each simulation was evolved for $10^5\,\tilde{P}$.

For null $\beta$, (blue lines on Figure \ref{fig:timeseries}) the system follows a familiar narrative: the test particle is captured into the 3:2 resonance, its eccentricity grows to the expected equilibrium, and the libration amplitude undergoes slow, long-term growth characteristic of overstability. As $\beta$ is increased to $\beta=2$ (gray lines on Figure \ref{fig:timeseries}; results for $\beta=1$ are qualitatively identical), however, the amplitude growth rate diminishes. For $\beta=4$ (green lines) the overstability is substantially weakened (leading to stable capture in a higher-index resonance), and for $\beta=8$ (red curve), the growth of the libration amplitude is essentially non-existent, signaling complete suppression of overstability. Results for $\beta=16$ are equivalent.

The numerical threshold $\beta_{\rm num}\simeq 8$ is within a factor of oder unity from the analytic condition derived in Section 2 (equation \ref{betacrit}), which predicts  $\beta_{\rm an}\simeq 13$. Indeed, the modest reduction factor $\chi\simeq 1/2$ arises from the full (non-WKB) dynamics of the resonant motion that accounts for contributions from dropped terms, such as $\ddot{\varpi}$. The simulations therefore verify that sufficiently sharp migration-rate gradients stabilize the resonance, even when the baseline configuration lies deep within the overstable regime. 

\section{Discussion}

The intriguing paradox of overstability is that while resonant configurations owe their very establishment to dissipation, the same dissipative forces can pull resonances apart. In this work, we have shown how a sharp dependence of the migration rate on semi-major axis -- as is expected in a migration trap -- can counteract this destabilizing tendency, allowing resonant librations to remain stable rather than grow.

 A useful way to understand how this quenching mechanism operates is to recall how overstability arises in the first place. Inside an interior resonance, the motion is governed by a conserved combination of semi-major axis and eccentricity, so that small librations necessarily couple the two: when $a$ dips slightly below its equilibrium value, $e$ rises, and vice versa (i.e., level-curves of the Tisserand parameter depicted on Figure \ref{fig:CRTBP}). Because the dissipative evolution of $a$ contains a term proportional to $e^{2}$, this correlation causes the orbit to experience semi-major axis decay more rapidly on the inward phase of each oscillation than on the outward phase. Over many cycles, the asymmetry amplifies the libration amplitude, driving the familiar overstable growth of the resonant angle. The semi-major-axis dependence introduced in this work offsets this effect. Near a convergence zone, the migration rate itself varies with $a$, such that inward excursions within the resonance experience a stronger outward push. This additional contribution counteracts the enhanced damping associated with the $e^{2}$ term, and can overtake it if the local torque gradient is sufficiently steep. In this regime, the net effect of a libration cycle is no longer to stretch the orbit away from the fixed point but to pull it back, and the resonance becomes linearly stable.

While the suppression mechanism we have identified follows directly from the dissipative structure of first-order interior resonances, the conditions under which it operates in practice -- i.e., $\beta \gtrsim (\tau_a/\tau_e)$ -- remain comparatively restrictive. A natural question therefore is: under what circumstances are real disks expected to satisfy the required torque-gradient condition? The answer is complicated by the fact that the most common migration trap -- the disk’s inner boundary -- is precisely where deviations from standard Type-I prescriptions are most pronounced. The surface-density profile there is sharp, possibly asymmetric, and partially mediated by MHD stresses (e.g., \citealt{2024MNRAS.528.2883Z}); accordingly, any analytic estimate obtained without a detailed numerical model should be viewed as being highly approximate.

Still, it is possible to extract the relevant orders of magnitude. Expressing the type-I disk torque as $\Gamma = \mathcal{C}\,\Gamma_0$ (where $\Gamma_0$ represents the reference torque, proportional to disk surface density, aspect ratio, etc., while the dimensionless multiplicative factor $\mathcal{C}$ captures the relative contributions of Lindblad and corotation components; see e.g., \citealt{2023A&A...670A.148P} and references therein), the migration-rate gradient may be written schematically as
\begin{align}
\beta &\sim -\frac{3}{2}\frac{d\ln|\Gamma_{\rm tot}|}{d\ln r}\sim \frac{3}{2}\,\frac{\ln\!\left(\mathcal{C}_{\rm e}/\mathcal{C}_{\rm g}\right)}{q\,(h/r)}  =\zeta  \left(\frac{h}{r}\right)^{-1}.
\end{align}
In the above expression, we have envisioned that the net torque reversal occurs across a radial width $\Delta r \sim q\,h$, with $h$ denoting the disk scale-height and $q$ of order unity. With the the corotation / horseshoe contribution near the edge ($\sim \mathcal{C}_{\rm e}$) being several times the differential Lindblad term dominating the power-law branch of the disk ($\sim \mathcal{C}_{\rm g}$) under favorable conditions \citep{2006ApJ...642..478M, 2009MNRAS.394.2283P}, we may take $\zeta$ as an order-unity constant that collects the numerical factors.

Independent of the exact torque regime, Type-I considerations require the ratio of eccentricity to migration timescales to scale as
\begin{align}
\frac{\tau_e}{\tau_a} \sim \xi\,\left(\frac{h}{r}\right)^{2}.
\end{align}
Numerical simulations of \citet{2021A&A...648A..69A} place $\xi$ in the range $\xi\simeq 5-15$. Recalling that our numerical experiments soften the analytic threshold for $\beta$ by a factor $\chi \lesssim 1$, we may recast our suppression criterion as a threshold aspect ratio:
\begin{align}
\frac{h}{r} \gtrsim \frac{\chi}{\zeta\,\xi}.
\end{align}
For representative values $\zeta\sim 1$, $\chi\sim 1/2$, and $\xi\sim 10$, this condition yields $h/r\gtrsim 0.05$, squarely within the range of typical protoplanetary disk parameters. Taken together, these estimates suggest that suppression of resonant overstability is a borderline phenomenon: it may operate in discs with sufficiently sharp torque transitions, but is unlikely to be universal, and will depend sensitively on the local thermodynamic, MHD, and saturation properties of the cavity edge. The phenomenon identified here thus opens an additional avenue by which disk-driven dynamics can modulate resonant survival.

The growing census of resonant planetary systems, nevertheless, offers a promising observational avenue to disentangle where and how commensurabilities are assembled. Intriguingly, hierarchical resonant pairs, where the outer planet is more massive than the inner, appear particularly suitable for such an analysis. In this vein, the recent work of \citet{BM26} showed that high-index resonances in such systems are most naturally established during the free-migration phase that extends over a substantial radial range. Conversely, the mechanism identified here points to the opposite end of the spectrum: resonant pairs that would otherwise be expected to succumb to overstability, yet survive in low-order commensurabilities, are natural candidates for capture near a torque reversal at the inner edge. For these systems, the observed resonance index and mass ratio may provide a direct handle on the sharpness of the local torque transition, i.e., the effective value of $\beta$ required to preserve the commensurability. As the sample of well-characterized resonant systems continues to expand, this line of reasoning may help turn resonance demographics into quantitative constraints on the structure of planet-disk torques at small stellocentric distances.



\appendix 

\section{Resonant detuning and sign of the apsidal precession}\label{app:signs}
In this appendix, let us briefly consider resonant detuning and sign of the apsidal precession from intuitive grounds. For an interior first-order resonance with an exterior perturber, the stable resonant geometry is the encounter-avoiding configuration in which successive conjunctions occur when the test particle is near periapse\footnote{A notable exception is the 2:1 resonance, which can also admit asymmetric stable equilibria once the eccentricity is sufficiently large; see, e.g., \citet{2006MNRAS.365.1160B}.}. This phasing minimizes repeated close approaches because the test particle is then at its greatest instantaneous separation from the perturber along its orbit (e.g., \citealt{1976ARA&A..14..215P}).

 At conjunction near periapse (where true anomaly $\varphi \sim 0$), the perturbing acceleration due to the exterior body is predominantly directed away from the central mass. Denoting the perturbing acceleration components in the osculating orbital frame as $\bar{F}=(R\,\hat{r},T\,\hat{\theta})$, we thus have $R>0$ and $T\sim0$. The sign of the resulting apsidal precession follows immediately from Gauss' perturbation equation \citep{1976AmJPh..44..944B},
\begin{align}
\dot{\varpi} = \frac{\sqrt{1-e^2}}{n\,a\,e}\left[-R\cos{\varphi} +T\,\sin{\varphi}\,\left(\frac{2+e\,\cos{\varphi}}{1+e\,\cos{\varphi}}\right)\right] \approx -\frac{R}{n \, a \, e} < 0. 
\label{eq:gauss_varpi}
\end{align}

The coherence of outward radial ``kicks'' applied at periapse explains why resonance drives \textit{retrograde} precession. Indeed, setting the acceleration to a fiducial estimate of $R\sim G\,m'/a^2$ yields
\begin{align}
\dot{\varpi} \sim -\frac{G\,m'}{n \, a^3 \, e}. 
\label{eq:gauss_varpi}
\end{align}
This reproduces equation (\ref{eqn5}) to within a factor of order unity. From equation (\ref{eqn4}), it then follows that the stable branch of the resonant fixed point corresponds to $\Delta = \dot{\varpi}/n < 0$, i.e., orbits that reside just wide of exact commensurability.

\section{Apsidal Acceleration and the Mass Dependence of Overstability}\label{app:B}
To see explicitly how the perturber mass re-enters the suppression criterion, it is useful to repeat the linear analysis without discarding $\ddot{\varpi}$. We keep all other assumptions and notation unchanged. Returning to Lagrange's equation (\ref{eqn5}), apsidal precession in the vicinity of the resonant equilibrium can be written as:
\[
\dot\varpi = n\,f\,\frac{m'}{M}\,\bigg(\frac{k-1}{k}\bigg)^{2/3}\,\frac{\cos(\phi_{\rm eq}+\psi)}{e_{\rm eq}+x}.
\]
Linearizing about the fixed point gives
\[
\delta\dot\varpi
=
\Delta_{\rm eq}\,\delta n
-\frac{n\Delta_{\rm eq}}{e_{\rm eq}}\,x
+\frac{\psi}{\tau_e},
\]
where we have used the fact that $\sin\phi_{\rm eq}=-1/(n\,\Delta_{\rm eq}\,\tau_e)$. 

Differentiating once more, we obtain terms proportional to $\dot{n}$, $\dot{x}$, and $\dot{\psi}$. Because $|\Delta_{\rm eq}| \ll (k-1)$, the $\dot{n}$ contribution to $\ddot\psi$ from $\ddot{\varpi}$ can be readily neglected. Moreover, the contribution arising from $\dot{x}$ can be written in terms of $x$ and $\psi$ using the linearized eccentricity evolution equation (\ref{eqn:dxdt}). We thus arrive back at the the same damped, driven harmonic oscillator
\[
\ddot\psi = -\omega_\mu^2\psi - \kappa_\mu\dot\psi - \gamma_\mu x,
\]
but with coefficients that now depend explicitly on $\Delta_{\rm eq}$ and $e_{\rm eq}$:
\[
\omega_\mu^2 = \omega^2+n^2\,\Delta_{\rm eq}^2,
\qquad
\kappa_\mu = \kappa+\frac{1}{\tau_e},
\qquad
\gamma_\mu = \gamma+\frac{n\,\Delta_{\rm eq}}{e_{\rm eq}\,\tau_e}.
\]

The WKB reduction now proceeds exactly as before, with the condition for suppression of overstability taking the form:
\[
\kappa_\mu > \gamma_\mu\,\eta\,\tau_e,
\]
or, equivalently,
\[
\frac{3\,\beta}{2\,\tau_a}+\frac{1}{\tau_e}
>
\left(\gamma+\frac{n\,\Delta_{\rm eq}}{e_{\rm eq}\,\tau_e}\right)\,\eta\,\tau_e.
\]
The mass of the perturber now enters the overstability condition explicitly through equilibrium expressions for resonant detuning and eccentricity, as well as the dimensionless coefficient, $\eta$.

\bibliographystyle{aasjournal}

\end{document}